\documentstyle[preprint,prl,aps,epsf]{revtex}
\draft

\tightenlines 
\title{Connection between Dispersive Transport and\\ 
       Statistics of Extreme Events}
\author{K.W. Kehr$^{a,\ast}$, K.P.N. Murthy$^{a,b}$, and H. Ambaye$^a$}
\address{$^a$ Institut f\"ur Festk\"orperforschung, Forschungszentrum
              J\"ulich GmbH,\\
              D-52425 J\"ulich, Germany\\
              $^b$ Materials Science Division,
              Indira Gandhi Centre for Atomic Research,\\ 
              Kalpakkam 603 102, 
              Tamil Nadu, India\\
              $\ast$ E.mail: k.w.kehr@fz-j\"ulich.de}
\begin{document}
\maketitle

\begin{abstract}

A length dependence of the effective mobility in the form of a power law, 
$B \sim L^{1-\frac{1} {\alpha}}$ is observed in dispersive transport 
in amorphous substances, with $0 < \alpha < 1$. We deduce this behavior
as a simple consequence of the statistical theory of extreme events.
We derive various quantities related to the largest value in samples 
of $n$ trials, for the exponential and power-law probability densities
of the individual events. 

\end{abstract}

\section{Introduction}

Dispersive transport in amorphous materials has been studied for
almost three decades now, and yet it continues to attract
considerable attention. In these studies, charge carriers are
created at one side of a slab of the material and transported
across to the other side. It has been realised that the  transport
phenomena in amorphous media cannot be described by standard concepts
like uniform drift and diffusional spreading, see 
\cite{pfisch,MW,schmidl,tiedje,baess} for exhaustive reviews 
on this subject and \cite{phystoday} for a popular account.
An important and striking observation is the
(apparent) dependence of the mobility on the thickness of the
material.  The general features of the mechanism of dispersive
transport in amorphous materials are well understood; the
process is subdiffusive and the delay in the transport 
occurs mainly due to trapping events in localized
centers of various energetic depths. A fairly large body of theory has
been developed to describe several aspects of the transport 
phenomena; already the early theoretical work of 
Shlesinger\cite{shle} and of 
Scher and Montroll\cite{schermon} could explain 
many of the experimentally observed phenomena, 
in particular the length-dependent mobility. In this note we
shall show that this length dependence of the mobility can be
easily understood as a simple consequence of the statistics of
extreme events.

The statistical theory of extreme events was formulated for
discrete random variables more than half a century ago by Fisher
and Tippett
\cite{FT} and Gnedenko \cite{gne}; it was extended and
popularized by Gumbel \cite{gum1,gum2}. Gumbel described various
applications that include statistics of extreme floods, droughts
and fracture of materials. In the field of condensed matter,
however,  the statistical theory of extreme events has  found
fewer applications.  The principal aim of the theory of extreme
events is to obtain the statistics of the largest value in a sample 
of $n$ independent realizations of a random variable. In particular 
the aim is to determine the asymptotic ($n \to \infty $) dependence
of the extreme values on the sample size $n$.
The random variable in the case of dispersive transport is the
residence time in the trapping centers. The question then can be
posed as follows.  Given the distribution of trap depths, how
does the largest trapping time increase with the number of
trapping centers; the latter quantity is determined by the
thickness of the material. (we shall  use length synonymously
with thickness). Together with the assumption that the extreme
events determine the behavior of the sum of the residence times,
the statistical theory of extreme events makes then a prediction
concerning the dependence of the largest residence time on
sample size or the thickness of the material.  Since mobility is
deduced from the sum of the transit times and this quantity is
given by the sum of the residence times, the length dependence
of the mobility would follow in a natural fashion, as we shall
show in this paper.

The paper is organized as follows. In section II the elements of
the statistical theory of extreme events that are necessary for
the later derivations are briefly described.  The application to
dispersive transport in disordered materials 
is discussed in section III. Detailed calculations of various quantities 
that are relevant for the characterization of the statistics of 
extreme events are made in section IV. The
paper closes with concluding remarks in section V.

\section{Elements of statistics of extreme events}

This section reviews standard material of the statistics of
extreme events
\cite {gum1,gum2} which is needed later on. Let $f(x)$ be the
probability density function (PDF) of a random variable $X$  and $F(x) =
\int^x dx'f(x')$ its  cumulative distribution function (CDF).
$F(x)$ is the probability that a particular realization of
$X$ has a value $\le x$.  Let $\Omega_n = \{ x_1 , x_2 , \cdots x_n \}$
be a set of $n$ independent realizations of $X$ sampled from the 
density $f(x)$. Let $x_{nl}$ denote the largest value of $X$ in the
set $\Omega_n$. The CDF of $x_{nl}$ is the same 
as the probability that all the values of $X$ in the set $\Omega_n$ are 
less than $x$, and hence is given by,  
\begin{equation}
\label{disn} \Phi_n(x) = F^n(x) \:.
\end{equation}
The probability density of the largest value of $X$ in the set $\Omega_n$ is
found by differentiation,
\begin{equation}
  \label{densn}
  \phi_n(x) = n\:F^{n-1}(x)\:f(x) \:.
\end{equation}

Given a value of $u$, let $g_u (\nu )$ denote the probability that the first 
$\nu -1$ values of $X$ are less than $u$ and the $\nu $- th 
value is greater than $u$. An expression for $g_u (\nu )$
can be readily written down as 
\begin{equation}\label{fpt}
g_u (\nu ) = F^{\nu -1}(u)[1-F(u)] \:.
\end{equation}
The mean value of $\nu$ is denoted by $n$ and is given by 
\begin{equation}
\label{mfpt}
n=\left\langle \nu \right\rangle =\sum_{\nu =1}^{\infty} \nu g_{u}(\nu)
                                 = {{1}\over{1-F(u)}} \:.
\end{equation}
The quantity $n$ is the mean number of steps required to exceed a given value
of $u$. 
The above relation can be `inverted' as follows. For a given $n$, considered now as
a parameter, let $u_{nl}$ denote the value of $u$ that obeys 
Eq. (\ref{mfpt}). Thus we get an implicit equation for $u_{nl}$ as,
\begin{equation}
\label{u_n} 
F(u_{nl})=1-\frac{1}{n} \:.
\end{equation}
Gumbel\cite{gum1} calls $u_{nl}$ as the 'expected' largest value of 
$X$ in a sample of $n$ independent realizations. Notice that $u_{nl}$ 
is {\it not} the mean of the extreme value in a sample of size $n$.
Hence $u_{nl}$ should strictly be viewed as a quantity defined by 
Eq. (\ref{u_n}).  
This simple expression for the `expected' largest value helps gain insight
into the asymptotic behaviour of the extreme values; we shall use this 
definition of 'expected' largest value to derive 
the power law dependence of the mobility in the next section.    

A simple example is provided by the exponential PDF $f(x)=\exp (-x)\:,\: x\ge 0$.
Solution of eq.(\ref{u_n}) with respect to $u_{nl}$ yields
\begin{equation}
\label{u_ne}
u_{nl} = \ln (n)\:, 
\end{equation}
i.e., a logarithmic increase of the expected largest value with sample size.

There are essentially three classes of behavior of the largest
value $u_{nl}$ for large $n$, leading to different asymptotic
forms of $\Phi_n(x)$ for large $n$
\cite {gum1,gum2}. The first class (I) is formed by probability densities 
$f(x)$ which decay at least exponentially for large $n$. The
second class (II) results from probability densities whose
moments diverge beyond a certain order.  The third class (III)
comprises probability densities where the values of $x$ are
bounded. We will encounter class I and class II behavior later,
depending on the physical quantity that is considered.

\section{Mobility in Dispersive Transport}

In a typical  experiment on dispersive transport, a slab of a finite
thickness $L$ (usually thin films of an amorphous substance) is 
coated with semitransparent metal electrodes. A constant voltage
is maintained across the slab. Charge carriers are created at
one surface at time $t=0$ by a laser pulse; the charge carriers
are drawn through the slab by the electric field. The current
$I(t)$ exhibits different behavior at short and at long times,
which is most clearly seen when plotted on a log-log graph.
Shlesinger \cite{shle} and Scher and Montroll \cite{schermon}
predicted
\begin{equation}
I(t)\; \sim \; \left\{ \begin{array}{l@{\qquad}l} t^{-1+\alpha }
& \mbox{$t \leq t_{tr}$} \\ t^{-1-\alpha } & \mbox{$t \gg
t_{tr}$}\:.  \end{array} \right.     \label{curr}
\end{equation}
where the parameter $0 < \alpha < 1$. Deviations from the ideal
behavior Eq.(\ref{curr}) are still a topic of current research
(see, e.g. \cite{baess,JK} and the references therein).  The physical
explanation of the behavior of $I(t)$ is by trapping of the
charge carriers in trapping centers with widely differing
depths. In the short-time regime, most of the charges are within
the slab, while at longer times the charges are extracted from
the slab.

A transit time $t_{tr}$ can be deduced from the crossover
between the short-time and long-time behavior. An effective
mobility is then defined by the ratio of effective velocity and
applied field $F$,
\begin{equation}
  \label{mob} B= \frac{L} {t_{tr}F} \:.
\end{equation}
In the multiple-trapping model, which is employed here, the
transit time $t_{tr}$ is the free transit time
$t_{free}$ plus  the sum of all dwell times $\tau_i$ in the
trapping centers. Similar arguments could be applied to
trap-controlled hopping models. The free transit time is given
by the mobility $B_0$, if there are no trapping events: $t_{free}
= L/(B_0F)$; it is usually short compared to the sum of all
dwell times. Hence to a good approximation
\begin{equation}
\label{ttr} t_{tr} \approx \sum_{i=1}^n \tau_i
\end{equation}
where $n$ is the number of trapping events. For a constant
trapping rate the number of trapping events is 
proportional to the thickness $L$ of the slab.

For broad distributions of dwell times $\tau_i$, the sum in
Eq.(\ref{ttr}) should be dominated by the largest dwell time.
With this assumption
\begin{equation}
  \label{est} t_{tr} \sim \tau_{largest} \:.
\end{equation}
The Arrhenius law is assumed for thermally activated processes,
\begin{equation}
  \label{Arrh} \tau_i = \tau_0\exp (\frac{E_i} {k_BT})
\end{equation}
where $E_i$ is the energy required for release from the trapping
center $i$. The largest dwell time is then determined by the
largest trapping energy. The probability density of the depths
of the trapping centers, {\sl i.e.}, of the energies necessary for
release is assumed to be exponential,
\begin{equation}
\label{DOS} 
f(E) =E_c^{-1}\exp\left( -\frac{E} {E_c}\right) \qquad \qquad
E \ge 0 \:.
\end{equation}
Note that $f(E)$ is normalized to unity in the interval       
$(0,\infty)$. 
It is easy to convert the probability density of the trapping energies into 
the probability density of the dwell times $\rho (\tau )$, using the Arrhenius
law Eq.(\ref{Arrh}). We have
\begin{equation}
  \label{rho}
  \rho (\tau ) = \frac{\alpha } {\tau_0} (\frac{\tau } {\tau_0})^{-(1+\alpha)}
\qquad \tau_0 \le \tau \ll \infty 
\end{equation}
with the parameter $\alpha = k_BT/E_c$. The probability density (\ref{rho}) is 
normalized, but already its first moment does not exist for $0 < \alpha < 1$, 
which is the parameter range of interest for dispersive transport. The cumulative 
distribution function of $\tau $ is given by 
\begin{equation}
  \label{cum2}
  P(\tau ) = 1 - (\frac{\tau } {\tau_0})^{-\alpha} \:.
\end{equation}
Equation (\ref{u_n}) for the expected largest value in a sample
of $n$ trials yields
\begin{equation}
  \label{tau_n} 
\tau_{nl} = \tau_0 n^{\frac{1} {\alpha}} \:.
\end{equation}
As already stated, the average number of trapping events $n$ is
proportional to the thickness of the experimental sample $L$. Hence
we expect, using the assumption that the transit time is
dominated by the largest dwell time
\begin{eqnarray}
  t_{tr} &\sim& L^{\frac{1}{\alpha}} \nonumber \\ B &\sim&
L^{1-\frac{1}{\alpha}} \:.
\end{eqnarray}
This is precisely the behavior of the mobility that has been
observed in experiments on dispersive transport, see for
instance \cite{schermon}.  Here we have derived this behavior
from the statistical theory of extreme events.

\section{Exponential and Power-law Probability Densities}

\subsection{Motivation}

Various questions arise with regard to the validity and the significance of the above 
result for the length dependence of the mobility. For instance, the meaning of the 
``expected largest value in a sample of $n$ trials'' is partially intuitive; i.e.
the precise meaning of the quantity that follows from Eq.(\ref{u_n}) is different
from what is suggested by this notion. Precisely defined quantities are the 
moments of the probability density function of the largest value in a sample of 
$n$ trials (if they exist) and the most probable value. Of course, complete information
is contained in the PDF itself. Fortunately, all quantities of interest can be 
derived exactly, if the underlying PDF for one event is exponential, or of power-law
form. This section will describe the results of these calculations, including a
numerical determination of the PDF of the dwell time.

\subsection{Exponential probability density for single events}

The basic quantity which determines the dwell times of particles in
the multiple-trapping model for dispersive transport is the trapping energy $E$.
It is a random quantity and the simplest, experimentally relevant, assumption 
is the exponential PDF, see Eq.(\ref{DOS}). The dimensionless form of this PDF
is $f(x) = \exp (-x)$ with the variable $x = E/E_C$ and the restriction
$0 \le x \le \infty$. The cumulative distribution function is then 
$F(x)=1-\exp (-x)$. The expected largest value in samples of $n$ trials has already
been given in eq.(\ref{u_ne}). The PDF for the largest value $x_{nl}$ in a sample of $n$
trials follows from Eq.(\ref{disn}) as
\begin{equation}
  \label{phi1}
  \varphi_n^{(1)}(x) = n (1-e^{-x})^{n-1}e^{-x} \:.
\end{equation}
The index $(1)$ shall indicate that this PDF is of type I in the classification of Gumbel
\cite{gum1,gum2}.

The moments of  $\varphi_n^{(1)}(x)$ can be calculated exactly.
The first moment, or mean value of $x_{nl}$ is defined as
\begin{equation}
\label{m1}
\langle x_{nl} \rangle = n \int_0 ^{\infty}dx\: x(1-e^{-x})^{n-1}e^{-x} \:.
\end{equation}
The evaluation of the integral is made in the Appendix. The result is 
\begin{equation}
  \label{r3}
  \langle x_n \rangle = \psi (n+1) - \psi (1) \:.
\end{equation}
The digamma function $\psi (z)$ is defined as \cite{AS}
\begin{equation}
  \label{dig}
  \psi (z) = {{d}\over{dz}} \ln \left( \Gamma (z)\right) \qquad {\rm and}
  \qquad          \xi (1) = - \gamma _E
\end{equation}
where $\gamma _E$ is the Euler constant. For integer $n$
\begin{equation}
  \label{intpsi}
  \psi (n+1) = -\gamma _E + \sum _{j=1} ^{n} {{1}\over{j}} \:.
\end{equation}
Thus
\begin{equation}
  \label{r4}
  \langle x_{nl} \rangle = \sum _{j=1} ^{n} {{1}\over{j}} \:.
\end{equation}
Euler's constant is asymptotically given by 
\begin{equation}
  \label{asE}
  \gamma _E = {}^{lim}_{n\to\infty} \left[ \sum_{j=1}^{n} {{1}\over{j}}
           - \ln (n) \right]
\end{equation}
Hence the first moment is given for large $n$ by
\begin{equation}
  \label{r5}
  \langle x_{nl} \rangle = \gamma _E + \ln (n) \:.
\end{equation}
Note that the first moment differs from the expected largest value of $x_{nl}$
by Euler's constant, cf. Eq.(\ref{u_ne}). Figure 1 contains the results on the 
first moment Eq.(\ref{r4}) and the expected largest value Eq.(\ref{u_ne}). Results 
on a numerical determination of the mean value $\langle x_{nl} \rangle$ have 
been included, to demonstrate that this quantity can also be accurately 
determined by numerical simulation.

The second moment $\langle x^2_{nl} \rangle$ can be calculated by the same 
technique as employed for the first moment, cf. the Appendix. The result for
the second moment is
\begin{equation}
\langle x^{2}_{nl}\rangle =
\left[ \psi (n+1) -\psi (1)\right] ^2 + \psi^{(1)}(1)-\psi^{(1)}(n+1)
\end{equation}
where $\psi^{(1)}(z)$ is the threegamma function defined by \cite{AS}
\begin{equation}
  \label{poly}
  \psi ^{(1)}(z)={{d^2}\over{dz^2}}\left( \ln \Gamma (z)\right) \:.
\end{equation}
The variance of the extreme value is then given by 
\begin{equation}
\langle x^{2}_{nl}\rangle - \langle x_{nl}\rangle^2  =
\psi^{(1)}(1)-\psi^{(1)}(n+1) \:.
\end{equation}
Therefore, the variance of the extreme value is 
\begin{equation}
\sigma ^2 = \sum_{k=1}^{n}{{1}\over{k^2}}=\xi (2)-\sum_{k=n+1}^{\infty}
{{1}\over{k^2}} \:.
\end{equation}
Note that $\xi (2) = \pi ^2 /6$. Hence the variance of the extreme
value approaches asymptotically this value, i.e., it is asymptocially 
independent of $n$. The significance of this fact has been stressed by
Gumbel \cite{gum1}. It is easy to derive the following bounds for the
variance:
\begin{equation}
{{\pi ^2}\over{6}} -\frac{1}{n} < \sigma ^2 < \frac{\pi ^2}{6} -\frac{1}{n+1} \:.
\end{equation}
The variance of the dimensionless extreme energy as a function of $n$ has been
included in Fig.1.

We return to the PDF for the largest value $x_{nl}$ in a sample of $n$
trials, which is generally given by Eq.(\ref{densn}) and for an
exponential PDF of single events by Eq.(\ref{phi1}). We define the most probable
value of the random variable $x_{nl}$ , {\sl i.e.}, the most probable extreme value, as 
the value of $x$ where the PDF $\varphi_n(x)$ has its maximum. Differentiation
of Eq.(\ref{densn}) yields the extreme value condition
\begin{equation}
  \label{extr}
  \varphi'_n (x)=n\left\{ F^{n-1}(x)f'(x)+(n-1)F^{n-2}(x)f(x)f(x)\right\} = 0 \:.
\end{equation}
This equation can be cast into a simple form,
\begin{equation}
  \label{ext2}
  {{n-1}\over{F(x)}}f(x)=-{{f'(x)}\over{f(x)}} \:.
\end{equation}
For an exponential PDF $f(x)=\exp (-x)$ the solution of this equation 
yields the most probable extreme value $x_{m.p.}$ as a function of the sample 
size, and is given by
\begin{equation}
  \label{mp1}
  x_{m.p.} = \ln (n) \:.
\end{equation}
Thus the most probable extreme value agrees with the expected extreme value in the 
case of the exponential PDF for the single events, cf. Eq.(\ref{u_ne}). 

\subsection{Power-law probability density for single events}

The quantity that determines the effective mobility is the sum of all dwell
times, which is assumed to be dominated by the largest dwell time. In this subsection
we direct the attention to the statistics of the largest dwell time.
The dwell time in a trapping center follows from the trapping energy via the 
Arrhenius relation Eq.(\ref{Arrh}). The PDF and the CDF for the single events
were already given in Eq.(\ref{rho}) and Eq.(\ref{cum2}), respectively. We will
use dimensionless variables $y =\tau /\tau_0$ henceforth. The PDF for the 
extreme value $y$ in a sample of $n$ events is then given by 
\begin{equation}
  \label{phi2}
  \varphi_n^{(2)}(y) = n\alpha (1-y^{-\alpha})^{n-1}y^{-(1+\alpha)}
\qquad 1 \le y < \infty 
\end{equation}
where the parameter $\alpha = k_BT/E_c$ and the relevant range is $0 < \alpha < 1$.
The index $(2)$ shall also indicate that the distribution of $y$ is of type II in the 
sense of Gumbel \cite{gum1,gum2}. 

The moments of $\varphi_n^{(2)}(y)$ can be calculated.
Let $\langle y_{nl}^{k}  \rangle$ denote the $k$th moment of the $y_{nl}$.
It is easily shown,
\begin{equation}
\langle y_{nl}^{k}\rangle = n \beta (1-{{k}\over{\alpha}},n)
\end{equation}
where $\beta (a,b) $ is the usual beta function.

>From the properties of the beta function it is clear that the $k$th moment of the 
extreme value exists only if $k$ is less than $\alpha$. 
Since in our problem, $\alpha$ is between zero and one, no integer 
moment $k \ge 1$ of the extreme value exists. 
Nevertheless, the expected extreme value (as defined by Gumbel) exists,
from which we have derived the length dependent mobility.

It is instructive to determine the most probable extreme value from the PDF 
Eq.(\ref{phi2}) for the largest value $y_{nl}$ in samples of n trials. The condition
for an extremum has been given in Eq.(\ref{extr}); evaluation with the PDF for
single events Eq.(\ref{rho}) and the CDF Eq.(\ref{cum2}) gives
\begin{equation}
\label{mp2}
y_{m.p.} = \left( {{ 1+\alpha n}\over{1+\alpha}}\right)^{1/\alpha} \:.
\end{equation} 
For large $n$, we can replace $1+\alpha n$ by $\alpha n$, and we get,
\begin{equation}
\label{mp3}
y_{m.p.}=\left({{\alpha}\over{1+\alpha}}\right)^{1/\alpha}n^{1/\alpha} \:.
\end{equation}

Another quantity, which is often used to characterize broad probability
densities, is the typical value. It is defined by
\begin{equation}
  \label{typ}
  \ln ( x_{typ}) = <\ln (x)> \:.
\end{equation}
where the brackets indicate a mean value taken with a general PDF.
The typical value of the PDF Eq.(\ref{phi2}) for the extreme value of $\tau$
is given by
\begin{equation}
  \label{typ2}
  \ln (y_{typ}) = \int_1^{\infty} dy \ln (y) n\alpha (1-y^{-\alpha})^{n-1}
   y^{-\alpha -1} \:.
\end{equation}
By the substitution $z=y^{-\alpha}$ this integral is transformed into
\begin{equation}
  \label{typ3}
  \ln (y_{typ}) = -\frac{n}{\alpha} \int_0^1 dz \ln (z) (1-z)^{n-1}
\end{equation}
The above integral can be evaluated exactly employing essentially the
technique described in the appendix (for the exponential PDF) and the
result is 
\begin{equation}
  \label{typ4}
  \ln (y_{typ}) = -\frac{1}{\alpha} [\psi (n+1) - \psi(1)] \:.
\end{equation}
Asymptotically, the typical extreme value behaves as 
\begin{equation}
  \label{typ5}
  y_{typ} \longrightarrow n^{\frac{1}{\alpha}} \exp (\frac{\gamma_E}{\alpha})  \:.
\end{equation}
That is, $y_{typ}$ is related to $<x_{nl}>$ by $y_{typ}=\exp(\frac{1}{\alpha}{<x_{nl}>})$.
We point out that the connection of the typical extreme value of
$y$ to the mean extreme value of $x$ is generally valid, if these  
two variables $x$ and $y$  are related by the exponential transformation
$y=\exp (x)$. Notice that the energy $E$ and the dwell time $\tau $ 
are related to each other by the Arrhenius law (\ref{Arrh}), which is
an exponential transformation. 

In Fig.2 we have plotted the expected largest value $\tau_{nl}/\tau_0$ of the
dimensionless dwell time, the most probable value according to eq.(\ref{typ4}),
and the typical value following from Eq.(\ref{typ4}), for $\alpha = 0.5$. 
In the case of the dwell time, the expected, the most probable, and the typical 
largest value are all proportional to $n^{\frac{1}{\alpha}}$, with different
$n$-independent factors for large $n$. It seems to be a general fact that
the quantities which characterize the largest value in samples
of $n$ trials, show similar behavior with respect to $n$, if they exist.

Although the PDF for the extreme value of $y$ in samples of $n$ trials is
exactly given by Eq.(\ref{phi2}),  the evaluation of this function is not
practical for large $n$. Therefore, we have determined this PDF for fixed
values of n by numerical simulations. We have generated random energies
according to the distribution Eq.(\ref{DOS}) with $E_c=1$ and calculated dwell
times using the Arrhenius law (\ref{Arrh}) with $\tau_0=1$ and $\alpha = 0.5$. 
The normalized
dwell times were sorted into bins containing $L$ values. The inverse of 
the lengths of the bins gives the PDF, when properly normalized. The 
result for $n=1024$ is given in Fig.3. The maximum of the distribution
agrees with the prediction Eq.(\ref{mp2}). The typical value is found 
right of the maximum, close but not identical to the median value. Similar
observations have been made for broad distributions previously, for instance
in Ref.\cite{KM}. 

Note that the PDF of the extreme value can have a different form, if it is
determined by different techniques. If the dwell times are binned into intervals
of logarithmically increasing intervals, the maximum of the resulting PDF
is found at a different location. This is due to the fact that by this
procedure another PDF is estimated, which is related to the one we used 
by the usual transformation with a Jacobian. The maximum of the latter
distribution is at $n^{\frac{1}{\alpha}}$ which is identical to the 
expected largest value as introduced by Gumbel. The message is that the
precise characterization of an extreme value depends on the PDF used for
this purpose.

A final problem is the justification of the replacement of the dwell times,
cf. Eq.(\ref{ttr}) by the largest dwell time, as was done in Eq.(\ref{est}).
In the subsequent derivation, the largest dwell time was identified with 
the expected largest time, however,  any quantitiy that characterizes an extreme
value can as well be used in the argument. The replacement can be justified by
extending the derivations given above to the most probable K$^{th}$ extreme 
value. Let $\varphi_{n,K}(x)$
denote the probability density function of the $K^{th}$ extreme value. A
formal expression for $\varphi_{n,K}(x)$ can be easily derived and is given
below:
\begin{equation}
\varphi_{n,K}(x)={{n!}\over{(n-K)!(K-1)!}}f(x)F^{n-K}(x)[1-F(x)]^{K-1}
\end{equation}
Note that if we set $K=1$, we recover the probability density
of the first extreme, which we have considered so far.
The condition for an extremal value is 
\begin{equation}
\varphi_{n,K}^{\prime}(x) = 0 \:. \label{ext5}
\end{equation}
Solution of this equation with respect to $x$ gives the most probable 
$K^{th}$ extreme value. 

The extremum condition Eq.(\ref{ext5}) can be solved for a power-law
PDF as given in Eq.(\ref{rho}). The result for the dimensionless variable
$y \equiv \tau /\tau_0 $ is
\begin{equation}
  \label{mpK}
  y_{m.p.}^{(K)} = \left(\frac{1+\alpha n}{1+\alpha K}\right)^{\frac{1}{\alpha}} \:.
\end{equation}
Let $r_K$ denote the ratio of the most probable 
$K^{th}$ extreme to that of the most probable (first) extreme. It is given by
\begin{equation}
  \label{rat}
  r_K= \left(\frac{1+\alpha }{1+\alpha K} \right)^{\frac{1}{\alpha}} \:.
\end{equation}
For $\alpha =0.5$ and $K=2$, we have $r=9/16$. The ratio becomes rapidly smaller
for larger $K$; it is already small for $K=2$ when $\alpha $ is small. The 
consequence is that the inclusion of the second, third, etc. extreme value
in the estimate of the dwell time would not change the asymptotic dependence
on the sample size $n$; only numerical factors would be modified.

\section{Concluding Remarks}

The apparent anomalies in the transport properties of charges in amorphous 
substances, which are summarized as 'dispersive transport',  are due to 
broad distributions of trapping times of the charge carriers. It is satisfying
that features like the length dependence of the effective mobility can already be
derived from the statistics of extreme events, the extreme events being the 
occurrence of particularly long trapping times. 

Although the length dependence of the effective mobility could already be deduced
from the notion of the ``expected largest value'' of the trapping time, we investigated
in detail various quantities in the context of the statistics of extreme events. 
Explicit expressions could be derived for the mean, the most probable, and 
the typical extreme values in the case of exponential or power-law probability densities 
for single events. We could also justify the use of the largest dwell time
to estimate the summary transit time by considering the second, third, etc.
extreme values. Their effect is to  modify numerical factors, but they do not 
alter the asymptotic dependence on the sample size.

All derivations  were made here for the ideal case of an exponential density
of energy levels. In practice deviations from the exponential density of states
are of interest. It turned out that many features of dispersive transport 
are also present for a Gaussian density of states \cite{baess}; if the 
temperature is sufficiently low. Hence it is of interest to extend the 
present derivations also to the case of other densities of states, for instance
to the Gaussian one. In the case of more complicated probability densites of the
single events, analytical derivations are either difficult or impossible. Hence one has 
to resort to numerical simulation to study the statistics of extreme events in those
cases.

Finally we point out that the observation of different behaviors of
distributions, depending on which of two exponentially related variables
are used, is rather general. For instance, de Arcangelis, Redner, and Coniglio
\cite{ARC} found broad distributions of voltage drops in random resistor
networks, whose moments are characterized by an infinite set of exponents.
To the contrary, the distribution of the logarithm of the voltage drops
behaves normally, in that the moments exhibit constant-gap scaling. Similar
behavior of the probability distributions of the cluster numbers in the
percolation problem was discussed by Stauffer and Coniglio \cite{SC}.
 
\vskip 1cm

Discussions with G. Sch\"utz are gratefully acknowledged.
KPNM is thankful to M.C. Valsakumar for numerous discussions on the
statistics of extreme events.

\section{Appendix: Evaluation of first moment of  $\varphi_n^{(1)}(x)$}

In the expression for the first moment Eq.(\ref{m1}) the substitution 
$y=\exp (-x)$ is made, 
\begin{equation}
  \label{m1s}
  \langle u_n \rangle = - \int _0 ^1 dy \: n ln (y) (1-y)^{n-1}  \:.
\end{equation}
The following trick \cite {mcv} is applied:
\begin{equation}
  \label{trick}
  \ln (y) = {}^{lim}_{\epsilon\to 0}
          {{y^{\epsilon} -1}\over{\epsilon}}
\end{equation}
We have
\begin{equation}
  \label{m1p}
  \langle u_n\rangle = {}^{lim}_{\epsilon\to 0}-n\int_0 ^1 dy\:
                      {{y^{\epsilon} -1}\over{\epsilon}}
                      (1-y)^{n-1} \:.
\end{equation}
The integral can now be performed with the result
\begin{equation}
  \label{r1}
  \langle u_n \rangle =-n {}^{lim.}_{\epsilon\to 0}
        {{\beta (1+\epsilon, n) - \beta (1,n)}\over{\epsilon}} 
\end{equation}
where $\beta (a,b) =\int _0 ^1 dx\: x^{a-1} (1-x)^{b-1} $. Hence 
we have
\begin{equation}
  \label{dbeta}
  \langle u_n \rangle =-n {{d}\over{dz}} \beta (z,n)\vert _{z=1} \:.
\end{equation}
The beta function is related to the gamma function (see \cite{AS}),
\begin{equation}
  \label{bg}
  \beta (a,b) = {{\Gamma (a) \Gamma (b)}\over{\Gamma (a+b)}} \:.
\end{equation}
Differentiating and observing that $\Gamma (n+1)=n \Gamma (n)$ and
$\Gamma (1) = 1$ we obtain
\begin{equation}
  \label{r2}
  \langle u_n \rangle = {{\Gamma ' (n+1)}\over{\Gamma (n+1)}}
                       - {{\Gamma ' (1)}\over{\Gamma (1)}} \:.
\end{equation}
Using the definition of the digamma function \cite{AS} this can be written as
\begin{equation}
  \label{r3a}
  \langle u_n \rangle = \xi (n+1) - \xi (1) \:.
\end{equation}

\newpage

\begin{center}
{\bf Figure Captions}
\end{center}

\vskip 0.5cm
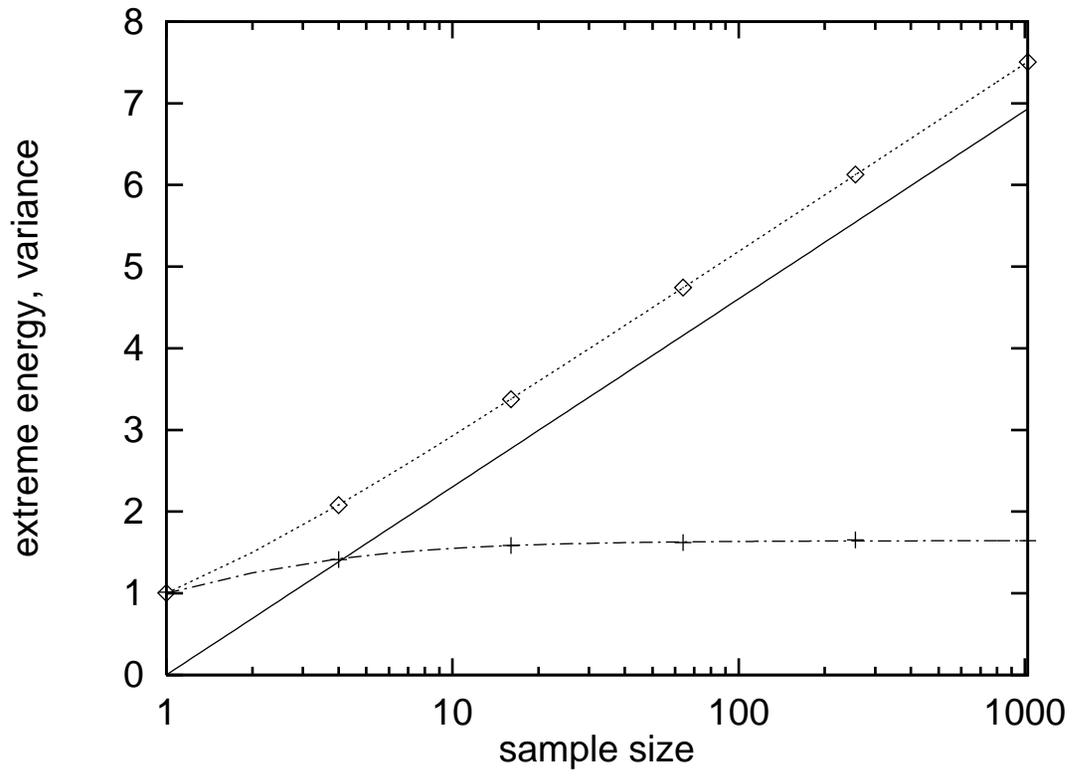
\noindent Fig. 1: Mean and variance of the dimensionless 
largest energy in samples of size
$n$, taken from an exponential PDF,
as functions of $n$. Full line: expected largest value $x_{nl}$, dotted line:
mean of largest value, dash-dotted line: variance of largest value. Symbols: 
$\Diamond$ :  mean largest value by simulations, 
$+$ : variance by simulations.

\vskip 0.5cm
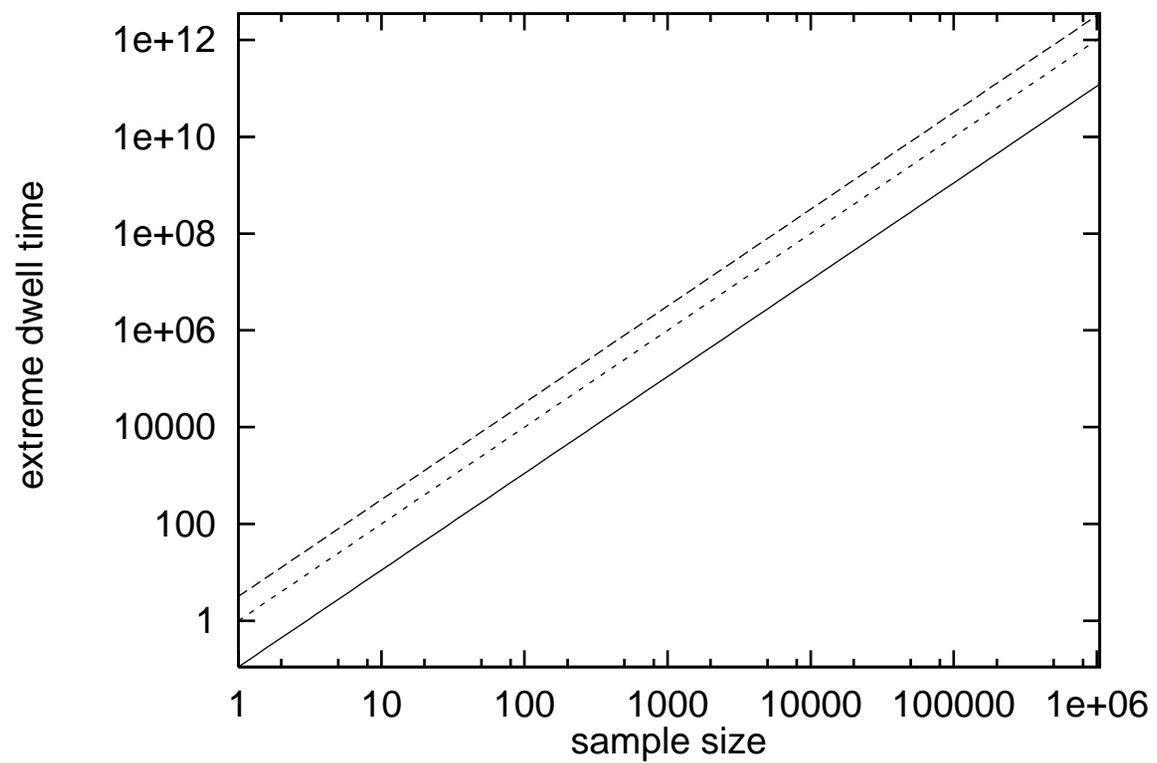
\noindent Fig. 2: Normalized extreme 
dwell times as  functions of the sample size $n$. 
Full line: most probable largest dwell time, short dashes: expected largest
dwell time, long dashes: typical largest dwell time.
 
\vskip 0.5cm
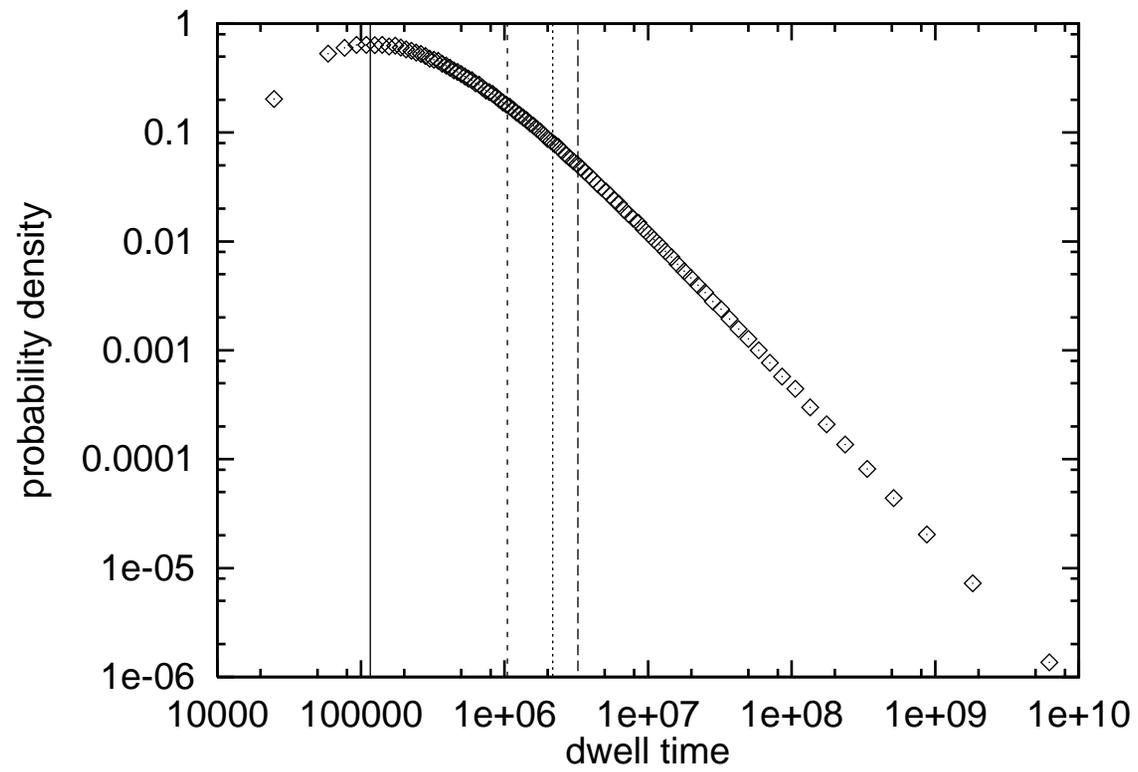
\noindent Fig. 3: Probability distribution function of the largest value of the
normalized dwell time in samples of size $n=1024$. The PDF was determined from
$10^5$ realizations. Indicated are: most probale larges value (full line),
expected largest value (short dashes), median (points), and typical largest
value (long dashes).

\end{document}